\documentclass[%
 aip,
 jmp,%
 amsmath,amssymb,
reprint,%
onecolumn]{revtex4-1}

\usepackage{graphicx}
\usepackage{dcolumn}
\usepackage{bm}
\usepackage{hyperref}
\usepackage{natbib}
\usepackage{amssymb}
\usepackage{amsmath}
\usepackage{stackengine}
\usepackage{pgfplots}
\usepackage{adjustbox}
\usepackage{makecell, multirow, tabularx}
\usepackage{mathtools}

\usepackage{float}
\usepackage{caption}
\usepackage{subcaption}
\pgfplotsset{compat=1.18}

\begin{document}

\title{Identifying Extreme Events in the Stock Market: A Topological Data Analysis}

\author{Anish Rai}
\email{anishrai412@gmail.com}
\affiliation{Department of Physics, National Institute of Technology Sikkim, Sikkim, India-737139.}

\author{Buddha Nath Sharma}
\email{bnsharma09@yahoo.com}
\affiliation{Department of Physics, National Institute of Technology Sikkim, Sikkim, India-737139.}

\author{Salam Rabindrajit Luwang}
\email{salamrabindrajit@gmail.com}
\affiliation{Department of Physics, National Institute of Technology Sikkim, Sikkim, India-737139.}

\author{Md.Nurujjaman}
\email{md.nurujjaman@nitsikkim.ac.in}
\affiliation{Department of Physics, National Institute of Technology Sikkim, Sikkim, India-737139.}
	
\author{Sushovan Majhi}
\email{s.majhi@email.gwu.edu}
\affiliation{Data Science Program, George Washington University, USA, 20052.}

\begin{abstract}

This paper employs Topological Data Analysis (TDA) to detect extreme events (EEs) in the stock market at a continental level. Previous approaches, which analyzed stock indices separately, could not detect EEs for multiple time series in one go. TDA provides a robust framework for such analysis and identifies the EEs during the crashes for different indices. The TDA analysis shows that $L^1$, $L^2$ norms and Wasserstein distance ($W_D$) of the world leading indices rise abruptly during the crashes, surpassing a threshold of $\mu+4*\sigma$ where $\mu$ and $\sigma$ are the mean and the standard deviation of norm or $W_D$, respectively. Our study identified the stock index crashes of the 2008 financial crisis and the COVID-19 pandemic across continents as EEs. Given that different sectors in an index behave differently, a sector-wise analysis was conducted during the COVID-19 pandemic for the Indian stock market. The sector-wise results show that after the occurrence of EE, we have observed strong crashes surpassing $\mu+2*\sigma$ for an extended period for the banking sector. While for the pharmaceutical sector, no significant spikes were noted. Hence, TDA also proves successful in identifying the duration of shocks after the occurrence of EEs. This also indicates that the Banking sector continued to face stress and remained volatile even after the crash. This study gives us the applicability of TDA as a powerful analytical tool to study EEs in various fields.

\end{abstract}
\maketitle

\begin{quotation}
Extreme events (EEs) are rare, unexpected occurrences that stand out significantly from normal happenings. These can include events like floods, heart attacks, power blackouts, and stock market crashes. For instance, the 2008 financial crisis and the COVID-19 pandemic caused massive stock market crashes worldwide, leading to significant financial losses for investors. Identifying and analyzing these extreme events is essential. One effective method for detecting EEs is topological data analysis (TDA), which can analyze multiple time-series data simultaneously. Using TDA-based norms and Wasserstein distance, we can identify EEs in stock markets across different continents. Additionally, we can conduct a sector-wise analysis to understand how various sectors were differently impacted during the COVID-19 pandemic based on their future outlook.

\end{quotation}

\section{Introduction}
\label{Introduction}

The stock market has witnessed various crashes starting from the great depression of 1929, the 1987 crash, the tech bubble of 2000, the 2008 financial crisis, and the latest COVID-19 pandemic~\cite{sornette2003critical,vandewalle1998crash,choi2012financial,mazur2021covid,rai2022statistical}. 
During these crashes, investors lost significant capital due to the panic sell-off caused by irrational decisions~\cite{chen2018panic,kaizoji2010bubbles,rabindrajit2024high}. 
However, the same crashes have provided opportunities for investors to make huge profits~\cite{kaizoji2010bubbles}. 
Due to the huge risk and opportunity, the study of crash dynamics in the stock market is significant.

An \emph{extreme event} (EE) is identified by the abrupt occurrence of unusual events~\cite{albeverio2006extreme,mahata2021characteristics}. EEs can happen in technological, social, and natural contexts~\cite{albeverio2006extreme}. They could originate naturally or due to human activity~\cite{raymond2020understanding}. Previously, floods, power blackouts, earthquakes, heart attacks, stock market crashes and upsurges were referred to as EEs.~\cite{mahata2021characteristics,korup2009natural,rai2023detection}. The study of EEs in different fields is important as the impact of these events is severe. There has been very limited work in the stock market in terms of the identification of EEs~\cite{mahata2021characteristics,rai2023detection}. 

Mahata et al.~\cite{mahata2021characteristics} recognized the stock price crash caused by the COVID-19 pandemic as an EE in different companies and indices. The analysis of each time series is carried out separately. The author applied the Empirical Mode Decomposition (EMD) based Hilbert--Huang transform (HHT) to identify EEs. The EMD-based HHT has been widely used to understand the stock market dynamics~\cite{mahata2020identification,mahata2020time}. Further, Rai et al.~\cite{rai2023detection} also identified the different factors leading to positive and negative EEs in different stocks using the EMD-based HHT. In these analyses, the authors have identified EE considering each stock price time series individually. 
However, when dealing with multiple time series, this method becomes impractical. Hence, we need a more convenient approach to analyze several time series simultaneously. That's where Topological Data Analysis (TDA) comes into play.

TDA attempts to solve data science problems using tools from algebraic topology and geometry~\cite{Carlsson2009TopologyAD,Dey_Wang_2022}. 
The idea behind employing TDA in our investigation is to extract the ``shape'' of a multi-dimensional time-series dataset. Using a fixed window of time, we first convert the slice of the time series into a Euclidean point cloud. Further, a topological signature---known as persistent homology---of the point cloud is computed. Now, as the slice moves along the time series, the evolution of the signature is then studied and analyzed~\cite{kulkarni2023investigation}. One of the key advantages of using persistent homology is its robustness under small, random perturbations of the data points. In contrast to other techniques that consider each time series separately, TDA allows us to analyze multiple time series at once in a complete and insightful manner---hence providing a more efficient and accurate identification of EEs. TDA has been successfully applied to problems in a growing number of fields, such as material sciences~\cite{kramar2013persistence,nakamura2015persistent}, 3D shape analysis~\cite{turner2014persistent}, multivariate time-series analysis~\cite{seversky2016time}, biology~\cite{yao2009topological},medicine~\cite{nicolau2011topology}, chemistry~\cite{lee2017quantifying}, network sensors~\cite{de2007homological}, early warning systems for flood~\cite{syed2021using}. Recently, TDA has also lent itself to analyzing the stock market crash dynamics. Some of the pivotal and relevant developments in this direction are discussed below.
\begin{itemize}
    \item The study of M. Gidea~\cite{gidea2017topology} detected early signs of a critical transition by tracking the topological changes when approaching a critical transition. The study analyzed the cross-correlation network of daily returns of the stocks under the Dow Jones Industrial Average index (DJIA). Likewise in Ref.~\cite{gidea2018topological}, the evolution of log returns of four major stock market indices---S\&P $500$, DJIA, NASDAQ, and Russell $2000$---was studied using persistent homology. In particular, it was observed that the $L^p$-norms of the persistence landscape showed strong growth in the vicinity of financial meltdowns. In addition to the four major US indices, Alejandro Aguilar et al.~\cite{aguilar2020topology} studied ten ETF sectors between January $2010$ and June $2020$ using algebraic topology and persistent homology. They studied the evolution of the $L^p$-norms over time to detect critical transitions in the daily log returns. The topological structures exhibited a significant change in behavior during the crash period.

\item Guo et al.~\cite{guo2020empirical}  analyzed the $L^1$ and $L^2$ norms during the 2008 crash in different regions. They observed a ``high-level period'' in norms before the crash. The market crash of 2020 in the Straits Times Index and Taiwan Capitalization Weighted Stock Index (TAIEX) was studied using TDA by Peter Tsung-Wen Yen et al.~\cite{yen2021using}. They studied the changes in the first three Betti numbers: $\beta_0$, $\beta_1$, and $\beta_2$ during market crashes. They observed that $\beta_0$ is small at the beginning of a market crash and increases as the market crash progresses. In a related work~\cite{yen2021understanding}, TDA was used to study the market crash in the Taiwan Stock Exchange (TWSE). The study found that the number of simplices grows gradually with increasing filtration parameters six months before the market crash, and grows swiftly near the crash.

\item Further, the persistence landscape was applied to perform enhanced indexing~\cite{goel2020topological}. Moreover, TDA was applied to cluster time-series models by their topological similarity and classification of the simulated dataset. This work provided evidence that stock price movements are sector-dependent~\cite {majumdar2020clustering}. In addition, a method that combines TDA with machine learning was used to understand the dynamics before a critical transition~\cite{gidea2020topological}. The authors applied Taken's theorem to estimate the persistence landscapes from Bitcoin time-series data. The so-called `Topological Tail Dependence Theory' was also proposed to bridge the gap between the mathematical theory of persistent homology and the financial stock market theory~\cite{souto2023topological}. The study found that incorporating persistent homology information systematically improves the forecasting accuracy of models, especially during turbulent periods. It also showed that the 2D Wasserstein distance is statistically significant for linear models. 

\item  Forty indices from forty countries/regions were analyzed through the persistence landscapes and their $L^p$-norms~\cite{guo2021analysis}. It was found that the connections between the stocks became closer and the norms were at the lower levels around the financial crises. In a related study, $100$ stocks from the Chinese Stock market were analyzed from $2013$ to $2020$~\cite{guo2022risk}. Through the strong correlation between the stocks, they found three turbulent periods in the time analyzed.
\end{itemize}.



Although a significant amount of study has already been carried out in the stock market using topology-inspired techniques, no work has yet identified the stock market crashes as an extreme event (EE) using TDA. Moreover, to the best of our knowledge, our analysis of the impact of different crises across continents and sectors is the first in the line. In our work, we have analyzed the crashes due to the 2008 financial crisis and COVID-19. These crashes are identified as EE using the TDA technique. During the COVID-19 pandemic, sectors were impacted differently based on their future outlook~\cite{mahata2021modeling}. Therefore, sector-wise analysis during the COVID-19 is important to understand their dynamics. Also, TDA is being applied for the first time for the identification of EEs. Mahata et al.~\cite{mahata2021characteristics} and Rai et al.~\cite{rai2023detection} have quantified EE in the stock market when the energy surpasses a particular threshold, i.e., $E_{\mu} +4*\sigma$. In our analysis, we have applied the Wasserstein distance $(W_D)$ and the $L^p$-norms of the persistent homology to detect the EE. The $W_D$ and the $L^p$-norms show an abrupt rise during stock market crashes. We have defined a threshold surpassing which the event is identified as an EE. To further verify the robustness of our results, we have applied the EMD-based HHT with the results obtained from TDA.

The structure of our paper is as follows.  Section~\ref{sec:methodology} explains the methodology employed in the paper and Section~\ref{our Approach} contains the TDA approach to analyzing time series. Section~\ref{Data} contains the data analyzed. In section~\ref{result} we have discussed our results and Section~\ref{con} contains the conclusion of the work.

\section{Methodology}
\label{sec:methodology}
\subsection{Topological Data Analysis}
\label{sec:TDA}

Topological data analysis (TDA) allows us to extract robust topological features from noisy datasets. One of the advantages of using TDA is its robustness, i.e., the output doesn't change under small perturbations~\cite{gidea2018topological}. In our work, we use Persistent homology which is a powerful tool in TDA, that provides a quantitative representation of how the shape and connectivity of data evolve as the resolution ($\varepsilon$) increases~\cite{gold2023algorithm}.

Sections~\ref{Ribs}--~\ref{PersistenceL} contain the definitions and notations from persistent homology used in the paper.

\subsubsection{Vietoris--Rips Complexes}
\label{Ribs}
To apply persistent homology, we start with a point cloud dataset, $X$, represented by a set of points in the $d$-dimensional Euclidean space $\mathbb{R}^d$. For $\varepsilon>0$, we associate a simplicial complex to the dataset using the Vietoris--Rips complex (also known as the Rips complex), denoted by $\mathcal{R}(X, \varepsilon)$. The Vietoris--Rips complex captures the connectivity of the data points at different $\varepsilon$. Although the Vietoris--Rips complex can be defined in a more general setting, we introduce the concept here in the context of Euclidean point clouds. A subset $\{x_0,x_1,\ldots,x_k\}$ of $X$ with $(k+1)$ points forms a $k$-simplex in $\mathcal{R}(X, \varepsilon)$ if all pairwise distances $\|x_i-x_j\|$ are at most $\varepsilon$. Roughly speaking, it represents a set of $(k+1)$ data points that are indistinguishable from each other at $\varepsilon$~\cite{gidea2018topological, souto2023topological, akingbade2024topological}.

The Vietoris--Rips complexes form a filtration, meaning that as we increase the value of $\varepsilon$, the simplicial complexes include more and more simplices~\cite{akingbade2024topological,chazal2021introduction}. For each complex in the filtration, we can compute its $k$-dimensional homology $H_k(\mathcal{R}(X, \varepsilon))$ with coefficients in some field. The homology groups capture the different dimensional topological features present in the dataset at each $\varepsilon$. For example, the $0$-dimensional homology group $H_0(\mathcal{R}(X, \varepsilon))$ corresponds to the connected components, the $1$-dimensional homology group $H_1(\mathcal{R}(X, \varepsilon))$ corresponds to the independent loops, the $2$-dimensional homology group $H_2(\mathcal{R}(X, \varepsilon))$ corresponds to the independent voids, and so on.

For two resolutions $0<\varepsilon_1\leq\varepsilon_2$, we can identify the birth and death of $k$-dimensional topological features by examining the homomorphism induced by natural inclusion~\cite{Dey_Wang_2022} $$H_k(\mathcal{R}(X, \varepsilon_1))\xhookrightarrow{\quad\iota_*\quad} H_k(\mathcal{R}(X, \varepsilon_2))$$ between the $k$-th homology groups. A non-trivial homology class in $H_k(\mathcal{R}(X, \varepsilon_1))$ may disappear in $H_k(\mathcal{R}(X, \varepsilon_2))$; likewise, a new homology class may appear in the latter that is not \emph{carried} by the inclusion from the former. 

\subsubsection{Persistence Diagrams}
\label{persistenceD}
The significance of the topological features is understood through their persistence. The features that persist for a long resolution range are called `significant' and the ones dying within a short resolution range are termed `noisy' features~\cite{guo2020empirical}. The summary of the birth and death of these features with change in resolution is preserved in a two-dimensional figure named \emph{persistence diagram}. It is a plot between the birth and death coordinates, consisting of two-dimensional points with their multiplicities.
Persistence diagram $\mathcal{P}_k$ corresponding to the $k$-dimensional homology consists of:
\begin{itemize}
    \item for each $k$-dimensional feature that is born at scale $b$ and dies at scale $d$, a point $p=p(b, d) \in \mathbb{R}^2$ together with its multiplicity $\mu$;
    \item $\mathcal{P}_k$  also contains all points in the positive diagonal of $\mathbb{R}^2$; these points represent all trivial homology generators that are born and instantly die at every level; each point on the diagonal has infinite multiplicity.
\end{itemize}

\begin{figure}[h!]
    \centering
    \begin{subfigure}[b]{0.45\textwidth}
        \includegraphics[width=\textwidth]{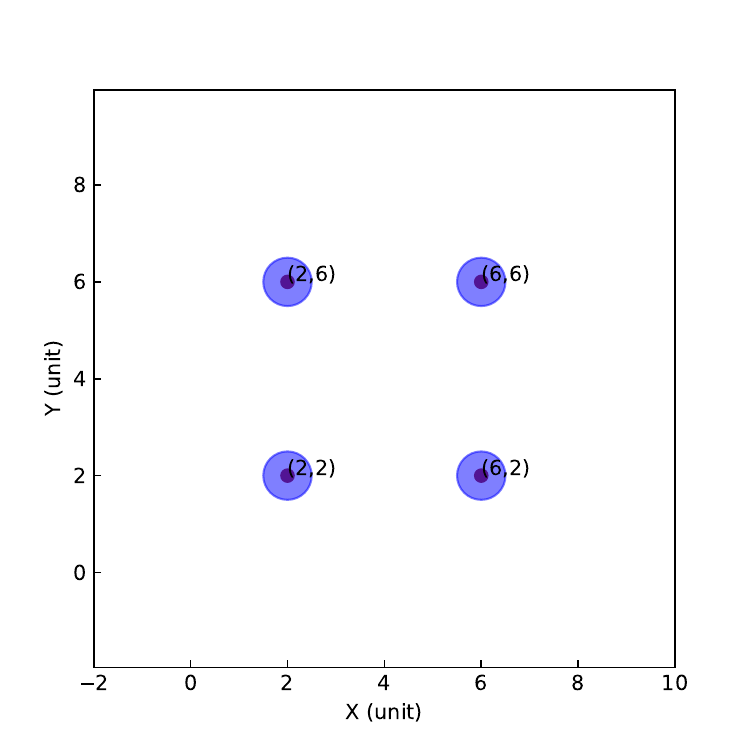}
        \caption{Rips Complex at $\varepsilon$=1}
        \label{fig:figure1_dummy}
    \end{subfigure}
    \hfill
    \begin{subfigure}[b]{0.45\textwidth}
        \includegraphics[width=\textwidth]{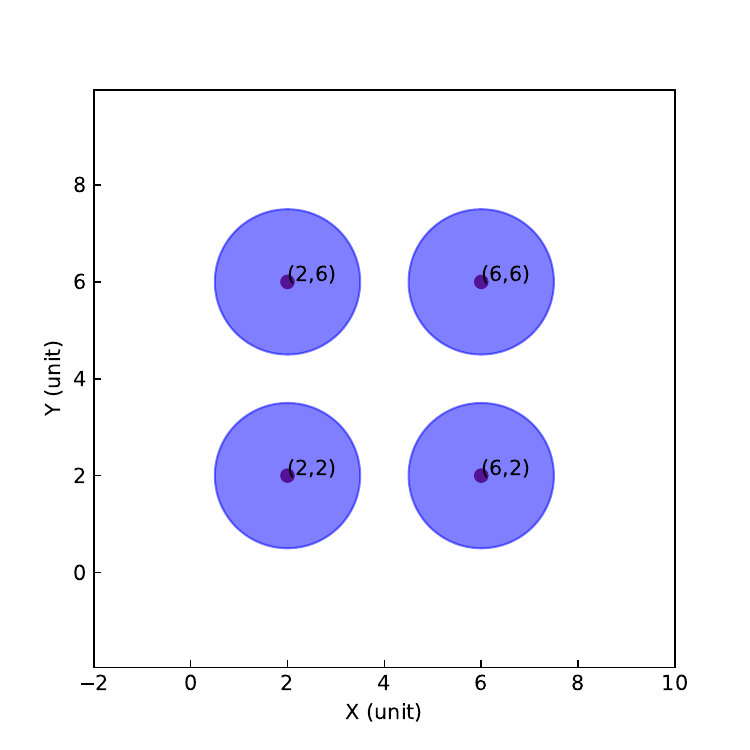}
        \caption{Rips Complex at $\varepsilon$=3}
        \label{fig:figure2_dummy}
    \end{subfigure}
    \medskip
    \begin{subfigure}[b]{0.45\textwidth}
        \includegraphics[width=\textwidth]{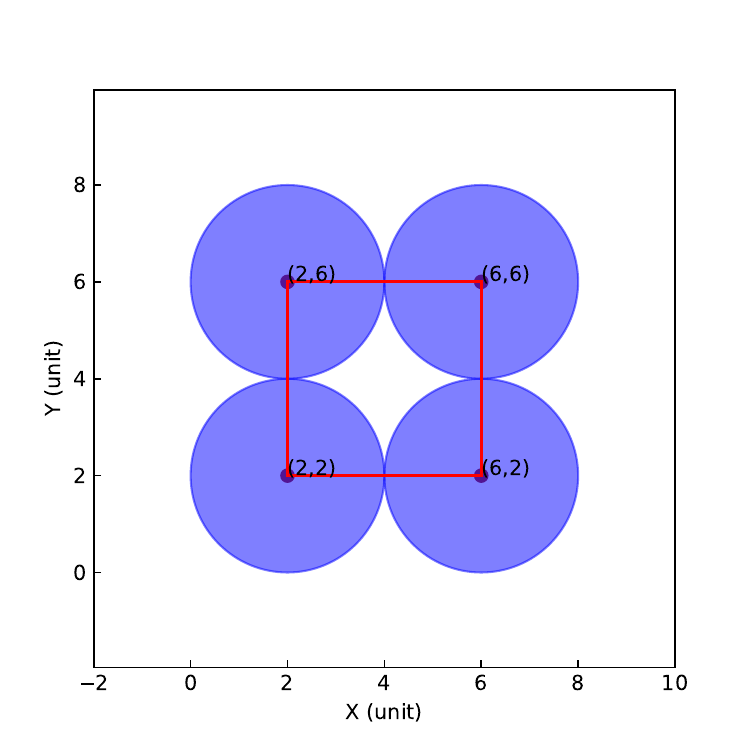}
        \caption{Rips Complex at $\varepsilon$=4}
        \label{fig:figure3_dummy}
    \end{subfigure}
    \hfill
    \begin{subfigure}[b]{0.45\textwidth}
        \includegraphics[width=\textwidth]{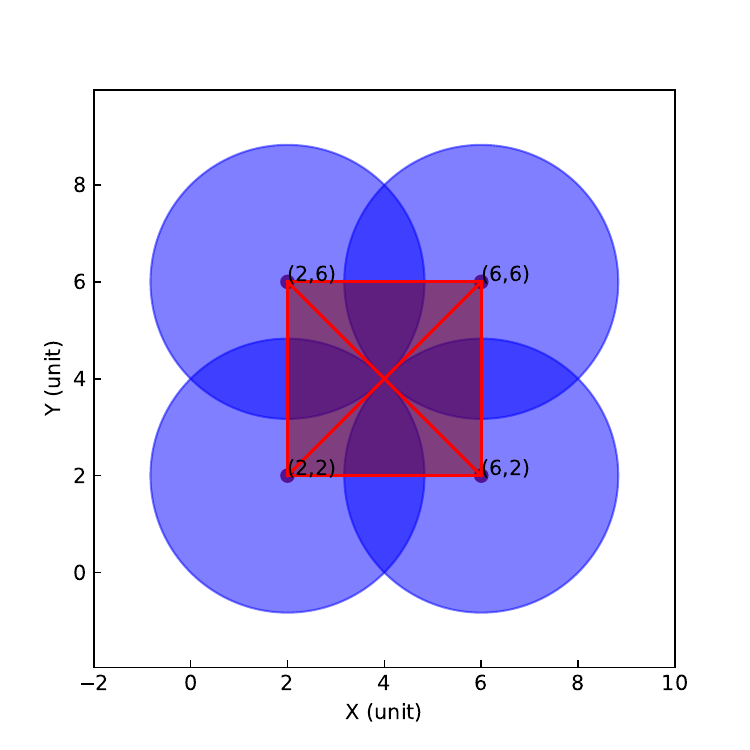}
        \caption{Rips Complex at $\varepsilon=4\sqrt{2}$}
        \label{fig:figure4_dummy}
    \end{subfigure}
    \caption{The plots represent the Vietoris--Rips complexes of a point cloud with four points at different resolutions($\varepsilon$). The disks in blue have been added as reference, to compare the mutual distances with the particular scale.} 
    \label{fig:overall_dummy} 
\end{figure}

\begin{figure}
    \centering
    \begin{subfigure}[b]{0.45\textwidth}
        \includegraphics[width=0.9\textwidth,height=5.0cm]{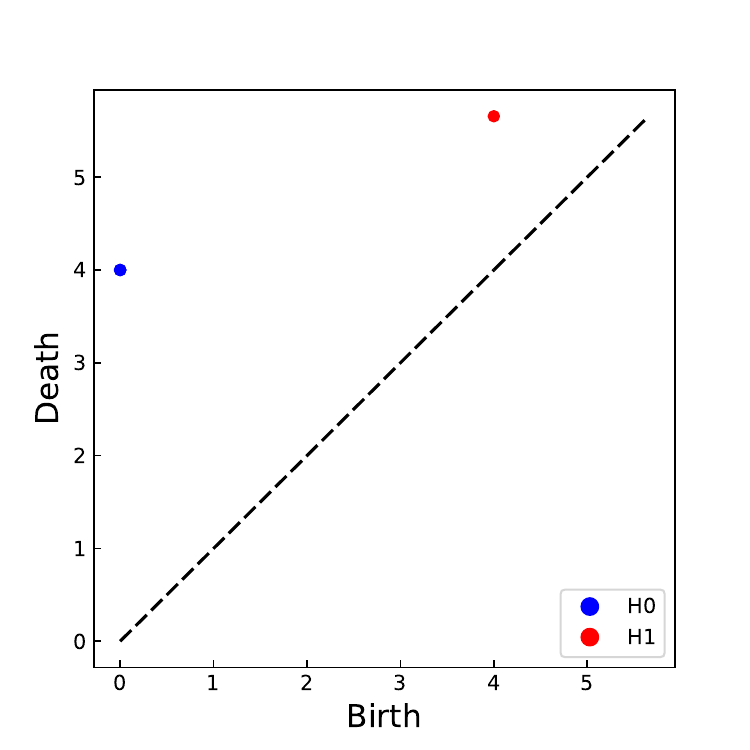} 
        \caption{
        The plot represents the Persistence diagram. The blue and red points correspond to the 0- and 1-dimensional holes respectively. Dashed lines ($---$) represent the positive diagonal.}
        \label{PD_dummy}
    \end{subfigure}
    \hfill
    \begin{subfigure}[b]{0.45\textwidth}
        \includegraphics[width=\textwidth,height=5.3cm]{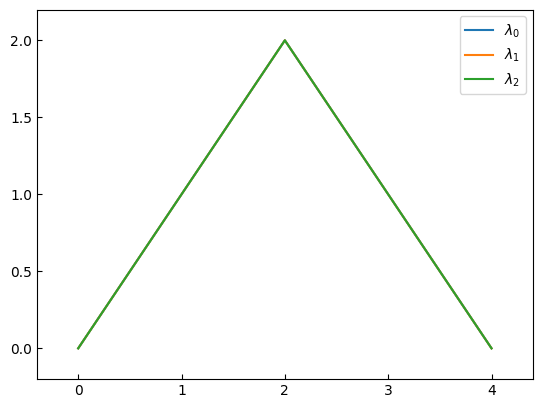}
        \caption{Plot represents the Persistence landscape for 0-dimensional homology group. $\lambda_0$, $\lambda_1$, $\lambda_2$ represents the first, second and third landscape respectively.}
        \label{PL_dummy}
    \end{subfigure}
    \caption{Plot (a) represents the Persistence diagram and plot (b) represents the Persistence landscape for the data.}
    \label{PDPL_dummy}
\end{figure}

To develop a better understanding of Vietoris--Rips complexes and persistence diagrams through visualization, let us consider the following set of four data points, $A = \{[2, 2], [2, 6],
[6, 2], [6, 6]\}$ in the two-dimensional Euclidean space. The Vietoris--Rips filtration for the above point-set are constructed at various $\varepsilon$ and shown in Fig.~\ref{fig:overall_dummy}. The points remain isolated for a range of $\varepsilon$ from $0$ to $3.9$ as shown in Figs.~\ref{fig:overall_dummy} (a) \& (b). We observe that at $\varepsilon = 4$, all the points get connected with each other except the diagonal pairs as shown in Fig.~\ref{fig:overall_dummy} (c). This marks the birth of a 1-dimensional hole. At $\varepsilon = 4\sqrt{2}$ the diagonal points also get connected forming a full polyhedron on the four points. The formation of the polyhedron indicates the death of the 1-dimensional hole. The complex at this resolution is shown in Fig.~\ref{fig:overall_dummy} (d). The summary of the birth and death of 0- and 1-dimensional holes is represented in the persistence diagram in Fig.~\ref{PDPL_dummy} (a). The Persistence landscape for the $0$-dimensional homology group obtained from the persistence diagram is shown in Fig.~\ref{PDPL_dummy} (b). 

\subsubsection{Wasserstein Distance}
\label{WassersteinD}
The space of persistence diagrams can be endowed with a metric structure. One commonly used metric is the degree $p$-Wasserstein distance~\cite{gidea2018topological}, denoted $W_D(\mathcal{P}_k^1, \mathcal{P}_k^2)$, where $\mathcal{P}_k^1$ and $\mathcal{P}_k^2$ are two persistence diagrams at dimension $k$. The distance $W_D(\mathcal{P}_k^1, \mathcal{P}_k^2)$ is defined as: 
\begin{equation}
\label{$W_D$}
W_D(\mathcal{P}_k^1, \mathcal{P}_k^2) = \min_{\phi: \mathcal{P}_k^1 \rightarrow \mathcal{P}_k^1} \left[ \sum_{x \in \mathcal{P}_k^1} \|x - \phi(x)\|_{\infty}^{p}\right]^{\frac{1}{p}}
\end{equation}
where $\phi$ is the bijective mapping from $\mathcal{P}_k^1$ to $\mathcal{P}_k^2$ and $\| \cdot\|_\infty $ is the sup norm. This metric measures the discrepancy between the two diagrams, taking into account the pairing of points between off-diagonal points and diagonal points. The Wasserstein distance provides a measure of similarity between persistence diagrams and is useful for comparing topological features across different datasets or analyzing the stability of features under perturbations.

A key advantage of persistence homology is its robustness under small perturbations. If the underlying data changes slightly, the corresponding persistence diagram only moves a small Wasserstein distance from the diagram of the original data~\cite{gidea2018topological,cohen2005stability}. This property makes persistence homology a valuable tool for analyzing complex systems.

\subsubsection{Persistence Landscapes and their Norms}
\label{PersistenceL}

The metric space formed by the persistence diagrams when endowed with Wasserstein distance $(\mathcal{P},\mathcal{W}_D)$, is not complete \cite{JMLR:v16:bubenik15a}. Hence, its statistical treatment becomes inappropriate. To achieve statistical analysis, persistence diagrams can be modified to persistence landscapes. It provides an alternative representation of persistence diagrams, which allows us to embed the space of persistence diagrams into the Banach space \(L^p(\mathbb{N} \times \mathbb{R})\), which is complete~\cite{JMLR:v16:bubenik15a}. To construct a persistence landscape from a persistence diagram \(\mathcal P_k\), we associate a sequence of functions \(\lambda = (\lambda_k)\), where \(\lambda_k: \mathbb{R} \rightarrow [0,1]\) is defined as follows:
\[
\lambda_k(x) = k - \max \left\{f_{(b_\alpha, d_\alpha)}(x) \mid (b_\alpha, d_\alpha) \in P_k\right\}.
\]
Here, \(f_{(b_\alpha, d_\alpha)}\) is a piecewise linear function defined as:

\[
f_{(b_\alpha, d_\alpha)}(x) = \begin{cases}x - b_\alpha, & \text{if } x \in\left(b_\alpha, \frac{b_\alpha + d_\alpha}{2}\right]; \\ -x + d_\alpha, & \text{if } x \in\left(\frac{b_\alpha + d_\alpha}{2}, d_\alpha\right); \\ 0, & \text{otherwise}.\end{cases}
\]
The persistence landscape \(\lambda\) is a sequence of functions that captures the ``heights'' of the piecewise linear functions \(f_{(b_\alpha, d_\alpha)}\) at each point \(x\) along the real line. The \(k\)-max denotes the \(k\)-th largest value of a function, and \(\lambda_k(x) = 0\) if the \(k\)-th largest value does not exist. Persistence landscapes form a subset of the Banach space \(L^p(\mathbb{N} \times \mathbb{R})\), which consists of sequences of functions \(\lambda = (\lambda_k)\), where \(\lambda_k: \mathbb{R} \rightarrow \mathbb{R}\) for \(k \geq 0\). This set has a vector space structure defined by \((\lambda^1 + \lambda^2)_k(x) = \lambda_k^1(x) + \lambda_k^2(x)\) and \((c \cdot \lambda)_k(x) = c \cdot \lambda_k(x)\) for all \(x \in \mathbb{R}\) and \(k \geq 1\). It becomes a Banach space when equipped with the norm \(\|\lambda\|_p\) given by~\cite{gidea2017topology}:
\begin{equation}
\label{norm}
\|\lambda\|_p = \left(\sum_{k=1}^{\infty}\left\|\lambda_k\right\|_p^p\right)^{1/p}
\end{equation}
where \(\|\cdot\|_p\) denotes the \(L^p\)-norm with respect to the Lebesgue measure on \(\mathbb{R}\). In this study, \(L^1\) and \(L^2\) norms are used.

\section{TDA Approach to Analyzing Multiple Time Series}
\label{our Approach}
\begin{figure}[h!]
\centering
\resizebox{0.55\textwidth}{!}{%
\tikzstyle{startstop} = [rectangle, rounded corners, minimum width=3cm, minimum height=1cm, text centered, draw=black, fill=red!50]
\tikzstyle{process} = [rectangle,rounded corners, minimum width=3cm, minimum height=1cm, text centered, draw=black, fill=brown!50]
\tikzstyle{decision} = [rectangle,rounded corners, minimum width=3cm, minimum height=1cm, text centered, draw=black, fill=green!50]
\tikzstyle{arrow} = [thick,->,>=stealth]

\begin{tikzpicture}[node distance=2cm, scale=0.05]

\node (start) [startstop] {1. Convert $n$ time-series into point could};
\node (slide) [process, below of=start] {2. Take sliding window of size $w$};
\node (construct) [process, below of=slide] {3. Construct Rips Complexes for each window};
\node (summarise) [process, below of=construct] {4. Construct Persistence Diagram for each window};
\node (decision) [decision, below of=summarise, yshift=-1.5cm] {5. Analyze Persistence Diagrams};
\node (Landscape)[process, below of=decision , right of=decision , node distance=2cm]{7 (a). Construct Persistence Landscape};
\node (Wasserstein)[process, below of =decision , left of=decision, xshift= -1.0cm, node distance=4cm]{6. Compute Wasserstein Distance};

\node (Lp)[process, below of=Landscape]{7 (b). Compute $L^p$-norm};

    \draw [->] (start) --  (slide);
    \draw [->] (slide) --  (construct);
    \draw [->] (construct) -- (summarise);
    \draw [->] (summarise) --  (decision);
    \draw [->] (decision) --  (Wasserstein);
    \draw [->] (decision) --  (Landscape);
    \draw [->] (Landscape) --  (Lp);
\end{tikzpicture}}
\caption{The flowchart represents our TDA approach to analyzing multiple time series.}
\label{fig:flowchart}
\end{figure}
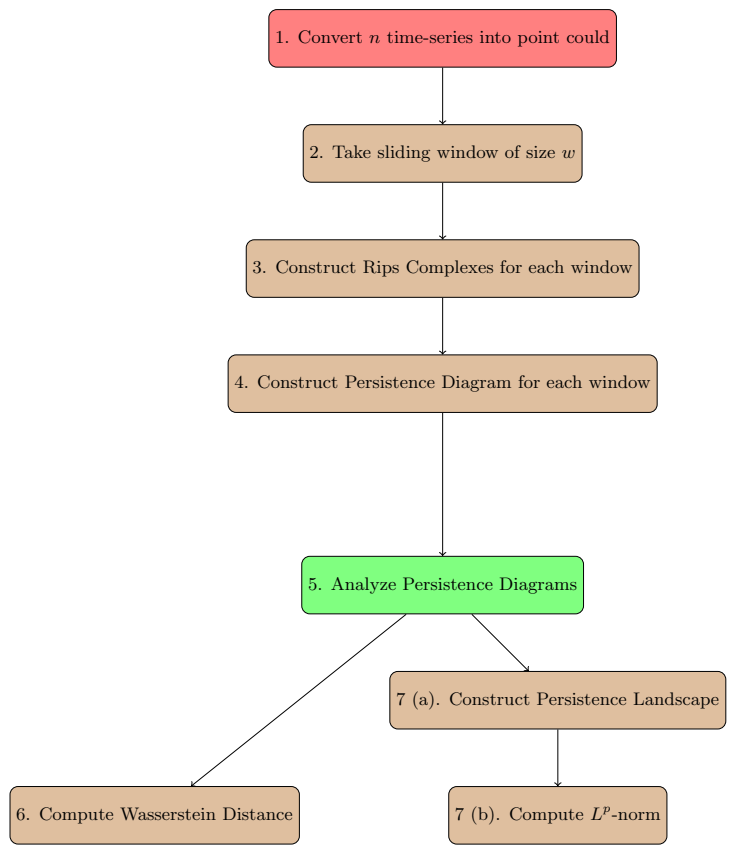

The sequence of computational steps used in TDA to obtain Wasserstein Distance ($W_D$) and $L^p$-norm from multiple time series is shown through a flowchart in Fig.~\ref{fig:flowchart}.

The detailed description of this flowchart is presented below:
\begin{enumerate}
\item Consider $n$ number of time series, each length $l$. For the $j$-th time interval, the $n$ time series form a point $(X_{1,j}, X_{2,j}, \ldots, X_{n,j})$ in $\mathbb{R}^n$. Considering all such points for $j=\{1,2,\ldots,l\}$, we get an $n$-dimensional Euclidean point cloud 
    $$D=\left\{\left(X_{1,j}, X_{2,j}, \ldots, X_{n,j}\right)\right\}_{j=1}^l\subset\mathbb{R}^n.$$ 

   \item We fix a sliding window of size $w$ to consider only a subset of the point cloud $D$ at a time. 
   
    \item For each position of the window, we construct the Vietoris--Rips filtration across various scales ($\varepsilon$).
    
    \item The birth and death of 0- and 1-dimensional holes are recorded in a persistence diagram for each window. We get a total of $l-w$ persistence diagrams for different window positions.

    \item The persistence diagrams obtained in the above step can be analyzed using various metrics like $W_D$ and $L^p$-norm. 
    
    \item The $W_D$ between consecutive persistence diagrams are calculated using Eq.~\ref{$W_D$}. $W_D$ compares the topological similarity in two persistence diagrams and its evolution can be studied to understand the dynamics of multiple time series considered.

 \item The persistence diagram can be further used in the following way:
    \begin{enumerate}
    \item Each persistence diagram is used to construct the persistence landscape. 

    \item The persistence landscape is used to calculate the $L^1$ \& $L^2$ norms by using Eq.~\ref{norm}. The norms are calculated for each persistence diagram and its variation with time is studied to comprehend the dynamics of multiple time-series considered.

    \end{enumerate}


\end{enumerate}

\section{Data Analyzed}
\label{Data}

We have carried out continent-wise analysis during the 2008 financial crisis and the COVID-19 pandemic. We selected four major indices from each continent—North-South America, Europe, Asia, and Oceania—to identify EEs, focusing on these highly developed regions. We have also analyzed the impact of COVID-19 on different sectors in India. The sectors include pharmaceuticals, banking, metals, automobiles, and fast-moving consumer goods (FMCG). We have taken the companies based on their respective Nifty sector indices. We have taken the daily closing price from January $1$, $2006$ to December $31$, $2010$ to analyze the $2008$ financial crisis and to analyze the crash due to COVID-19 we have taken the data from January $1$, $2019$ to December $31$, $2022$. The time duration has been chosen as it captures the whole crash period. The data have been obtained from Yahoo Finance Website~\cite{Yahoo}. The subsequent sections present the results of our analysis.

\section{Results and Discussion}
\label{result}

This section shows the result of the identification of continent-wise extreme events (EEs) during the 2008 financial crisis and the COVID-19 pandemic using TDA. It allows the identification of EEs from multiple stock time series at 
once. Also, a sector-wise impact of the COVID-19 pandemic is analyzed in the Indian stock market.

\begin{figure}[h!]
    \centering
    \begin{subfigure}[b]{0.45\textwidth}
        \includegraphics[width=\textwidth,height=5.0cm]{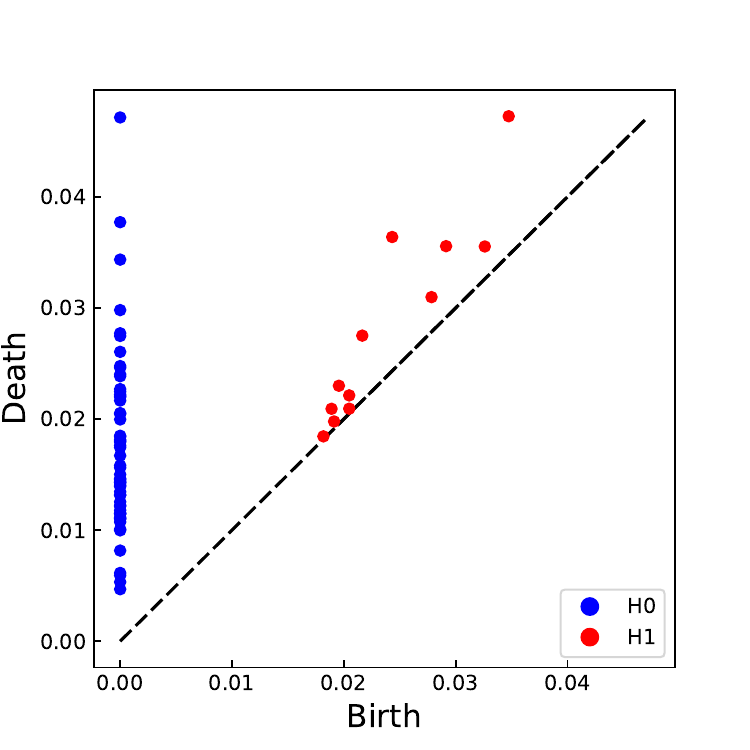}
        \caption{Persistence diagram for the point cloud. The blue and red points correspond to the 0- and 1-dimensional holes respectively. Dashed lines $---$ represent the positive diagonal.}
        \label{PD1}
    \end{subfigure}
    \hfill
    \begin{subfigure}[b]{0.45\textwidth}
        \includegraphics[width=\textwidth,height=5.3cm]{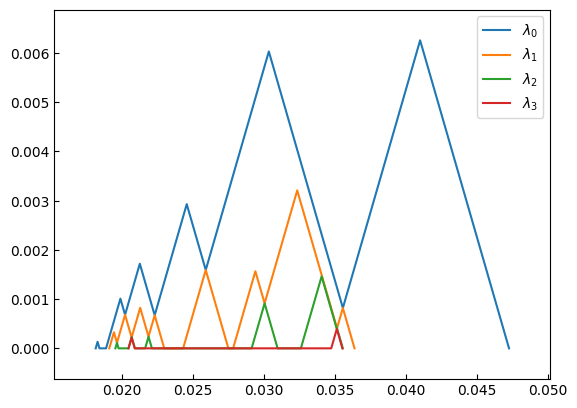}
        \caption{The persistence landscape for 1-dimensional homology group. $\lambda_0$, $\lambda_1$, $\lambda_2$, $\lambda_3$ represents the first, second, third and fourth landscape respectively.}
        \label{PL1}
    \end{subfigure}
    \caption{Plot (a) represents the Persistence diagram and plot (b) represents the Persistence landscape for point cloud obtained for North-South America.}
    \label{PDPL}
\end{figure}

\subsection{Obtaining point cloud from stock price time-series}
\label{identification}

In order to apply TDA in the stock market, we construct a point cloud from the stock price time series. We calculate the log-return of $n$ stock price time-series of length $l$ as $R_{i,j}\coloneqq\ln{\frac{P_{i,j}}{P_{i,j-1}}}$, where $P_{m,k}$ denotes the closing price of the index $m=\{1,\ldots,n\}$ on day $k \in \{1,\ldots,l\}$.

By taking the log-return of all time-series on the $j$-th day, i.e., $(R_{1,j}, R_{2,j}, \ldots,R_{n,j})$, a point is formed in $\mathbb{R}^n$. An $n$-dimensional Euclidean point cloud 
$\{(R_{1,j}, R_{2,j}, \ldots, R_{n,j})\}_{j=1}^l$
is obtained by taking all such points for $j=\{1,2,\ldots,l\}$, After obtaining the point cloud, we have followed the steps discussed in Section~\ref{our Approach} to estimate the $W_D$ and $L^p$-norms.

\begin{figure}[h!]
    \centering
    \includegraphics[angle=0, width=15cm, height=10cm]{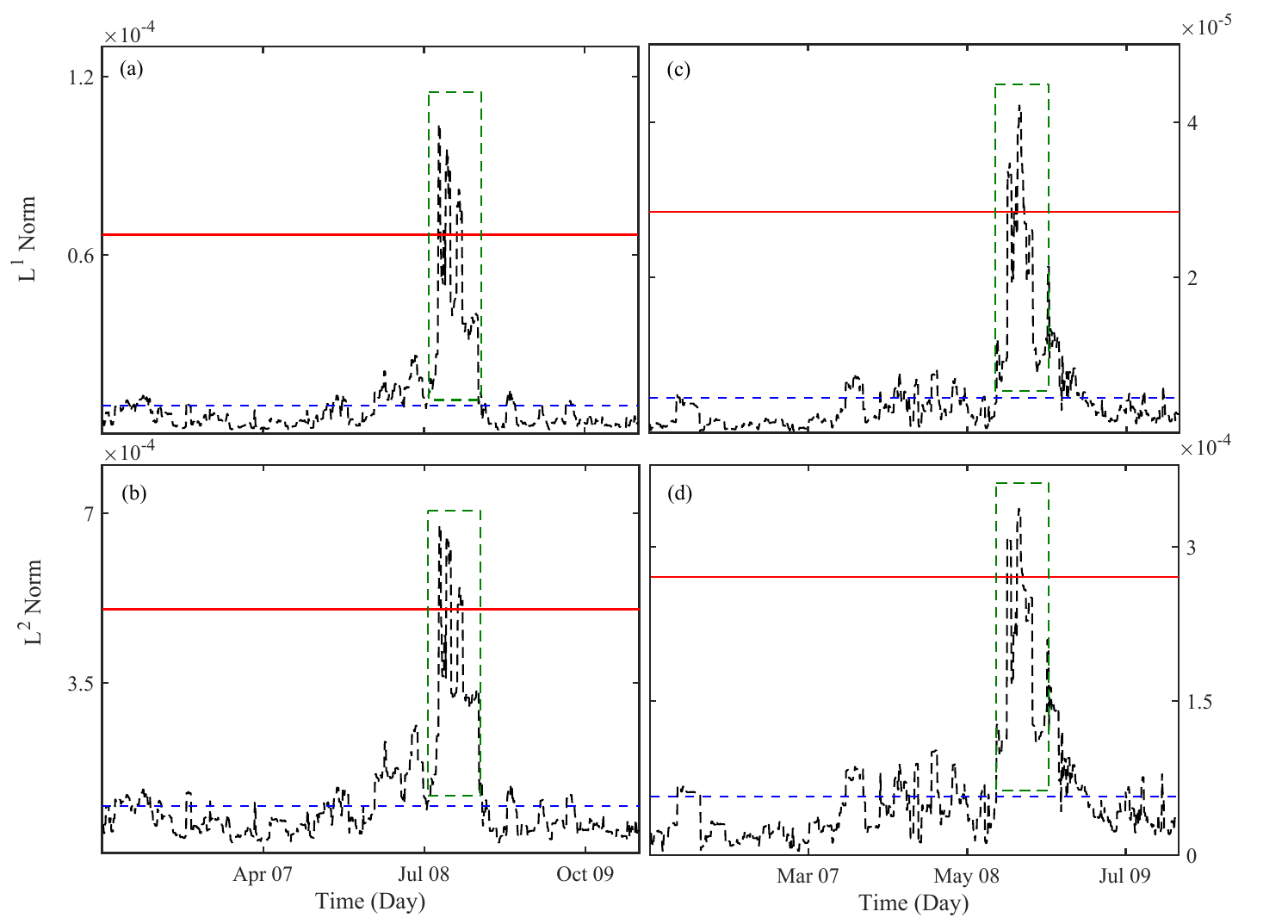}
    \caption{Plot (a) and (b) represent the $L^1$ and $L^2$ norms of North-South America and plot (c) and (d) represent the $L^1$ and $L^2$ norms of Europe continent, respectively. The abrupt rise in the norm values above the threshold during 2008 confirms the occurrence of EE.}
    \label{fig:USEurope_L}
\end{figure}

\subsection{EE due to the 2008 Financial crisis}

To identify EE, we define a threshold on $W_D$ and $L^p$-norms above which the event is considered as EE. The threshold applied is $\mu+4*\sigma$, where $\mu$ and $\sigma$ are the mean and standard deviation of the $W_D$ or norm values.

Figures~\ref{PDPL} (a) \& (b) represent the Persistence diagram and its $1$-dimensional Persistence landscape of North-South America for a window size of 60. We have estimated the $L^1$ \& $L^2$ norms of the persistence landscapes during the 2008 financial crisis. Figs.~\ref{fig:USEurope_L} (a) \& (b) represent the $L^1$ and $L^2$ norms, respectively, of North-South America. The black dashed lines represent the norm values, and the horizontal red and blue lines represent the threshold and the average value of the norms, respectively. We have defined the threshold at $\mu_N + 4*\sigma$ where $\mu_N$ and $\sigma$ are the mean and the standard deviation of the norm values, respectively. There is a clear spike in the norm values during late 2008, shown by the green rectangular box, which corresponds with the $2008$ financial crisis. Hence, we conclude that TDA successfully detected the crash during the $2008$ financial crisis in North-South America. Also, as the spike in the norm values exceeds our predefined threshold, the crash in America due to the $2008$ financial crisis can be considered an EE.

Similarly, Figs.~\ref{fig:USEurope_L} (c) \& (d) represent the $L^1$ and $L^2$ norms, respectively, of Europe. A clear spike in both the norm values is distinctly visible and corresponds with the crash due to the $2008$ financial crisis. The norm values during this period surpass the threshold value which is marked by the red horizontal line. This clearly shows that the crash in Europe due to the 2008 financial crisis is an EE. We have carried out our analysis for Asia and Oceania and we have obtained similar results showing EE during the $2008$ financial crisis.

\begin{figure}[h!]
    \centering
    \includegraphics[angle=0, width=15cm, height=5cm]{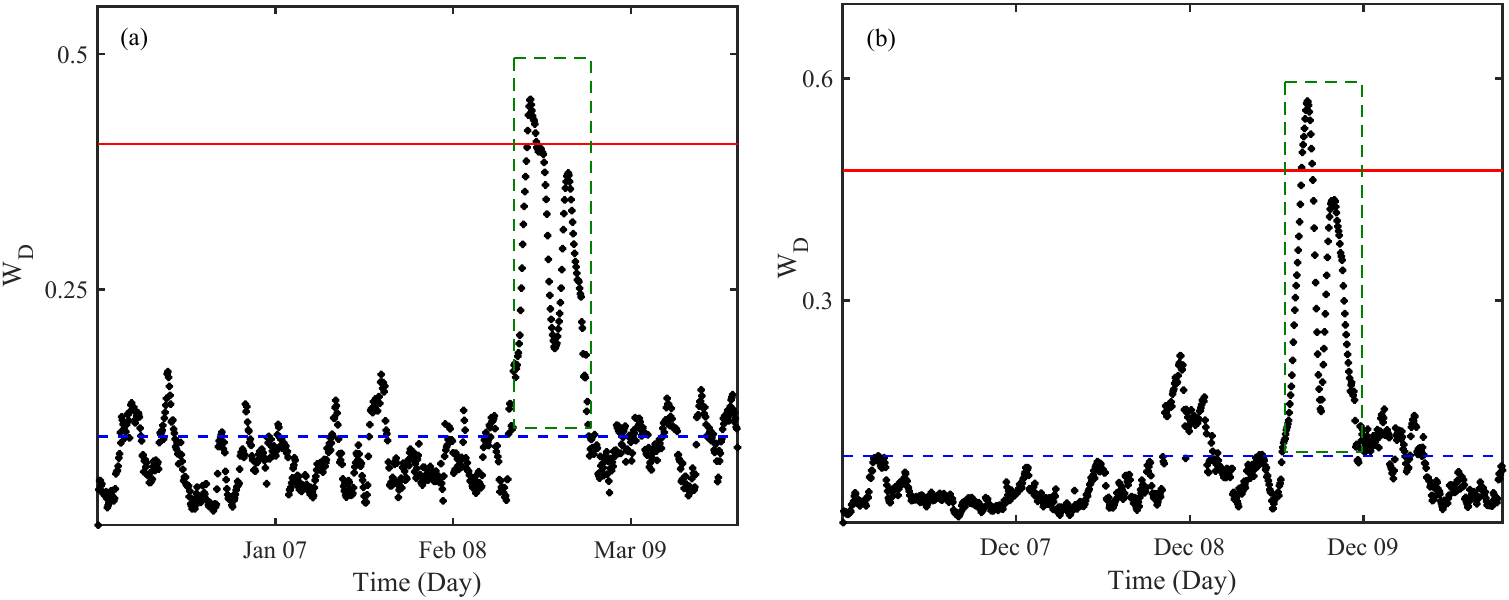}
    \caption{Plot (a) and (b) represent the Wasserstein distance ($W_D$) of North-South America and Europe, respectively from 2006 to 2010. The spike in the $W_D$ above the threshold value coincides with the 2008 financial crisis}
    \label{fig:UE_W}
\end{figure}
We also estimate the Wasserstein distance ($W_D$) to analyze the market dynamics during the 2008 financial crisis. Figs.~\ref{fig:UE_W} (a) \& (b) represent the $W_D$ of North-South America and Europe, respectively. The blue and red lines represent the average value of the Wasserstein distance ($\mu_W$) and the threshold value. The threshold value is defined as $\mu_W + 4*\sigma$, where $\sigma$ is the standard deviation of the $W_D$. The abrupt rise in the $W_D$ values surpassing the threshold can be regarded as an EE. The threshold of $L^1$, $L^2$ norms, and $W_D$ for different continents during the 2008 financial crisis is listed in Table~\ref{Table 1}. This identification of EE using TDA is consistent with the existing literature done in Ref.~\cite{rai2023detection}.

\renewcommand{\arraystretch}{1.5} 
\begin{table}[h!]
\centering
\begin{tabular}{ |p{3cm}|p{2cm}|p{2cm}|p{1cm}|p{2cm}|p{2cm}|p{1cm}|  }
 \hline
  \multirow{3}{*}{Continents}& \multicolumn{3}{|c|}{2008 Financial crisis} &\multicolumn{3}{|c|}{COVID-19} \\
\cline{2-7}
& $L^1$-norm    &$L^2$-norm& $W_D$& $L^1$-norm    &$L^2$-norm& $W_D$\\
 \hline
America & $6.68\times 10^{-5}$ &$5.02\times 10^{-4}$ &0.41 &$4.27\times 10^{-5}$ &$3.65\times 10^{-4}$ &0.47 \\
 \hline
Asia & $1.00\times 10^{-4}$ &$6.33\times 10^{-4}$ &0.65 &$4.27\times 10^{-5}$ &$3.70\times 10^{-4}$ &0.38 \\
 \hline
Europe & $2.84\times 10^{-5}$ &$2.71\times 10^{-4}$ &0.48 &$9.66\times 10^{-6}$ &$1.26\times 10^{-4}$ &0.43 \\
 \hline
Oceania & $3.00\times 10^{-6}$ &$5.41\times 10^{-5}$ &0.37 &$1.17\times 10^{-5}$ &$1.36\times 10^{-4}$ &0.34 \\
 \hline
\end{tabular}
\caption{Table represents the threshold of $L^1$ norm, $L^2$ norm, and $W_D$ of different continents during the crash due to the $2008$ financial crisis and  COVID-19 pandemic.}
\label{Table 1}
\end{table}
\renewcommand{\arraystretch}{1} 

\begin{figure}[h!]
    \centering
    \includegraphics[angle=0, width=15cm, height=10cm]{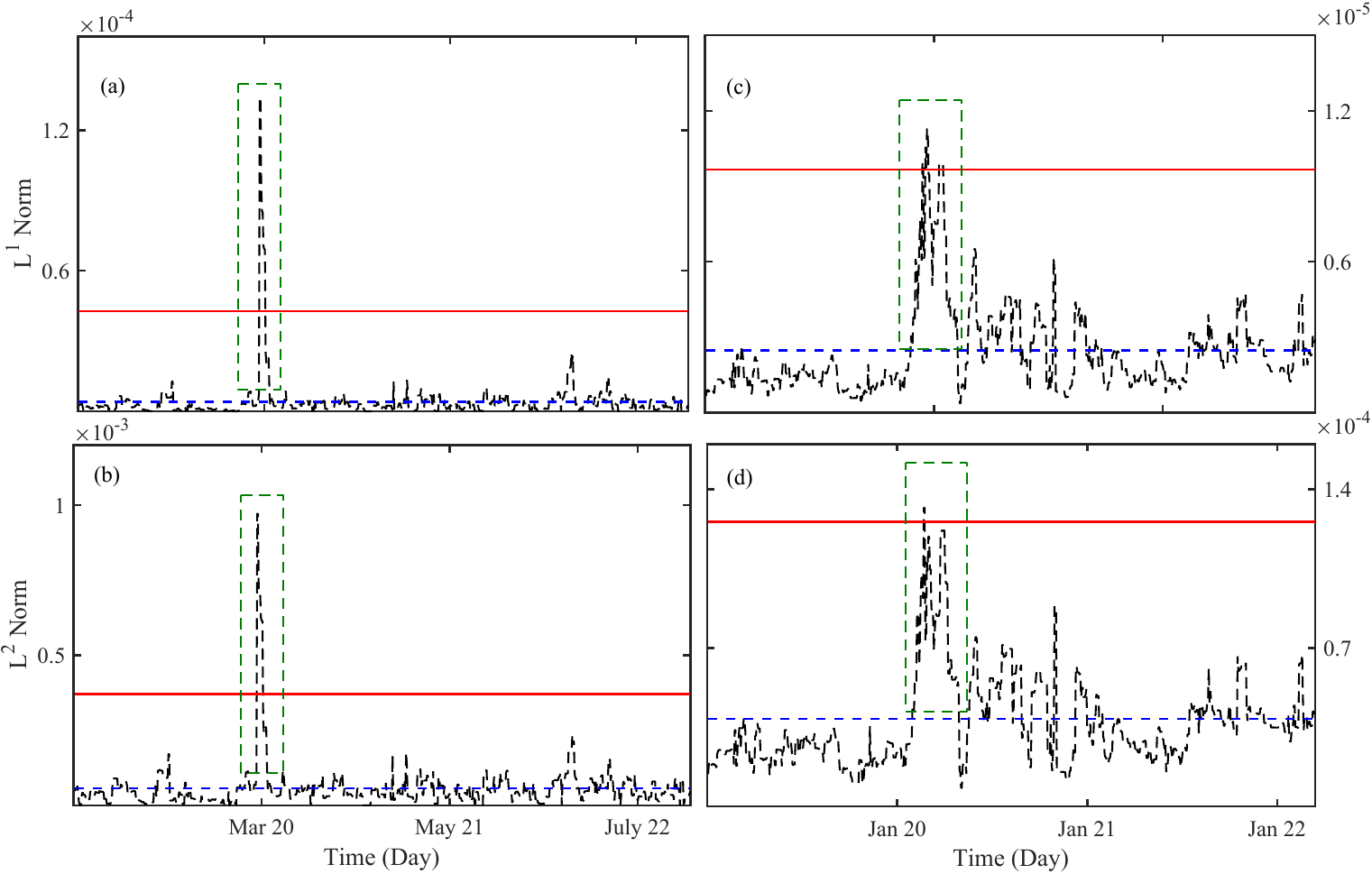}
    \caption{Plot (a) and (b) represent the $L^1$ and $L^2$ norms of Asia and plot (c) and (d) represent the $L^1$ and $L^2$ norms of Europe continent, respectively. The abrupt rise in norm values above the threshold confirms the EE due to COVID-19.}
    \label{fig:AsiaEurope_COVID}
\end{figure}

\subsection{EE due to COVID-19 pandemic}

\begin{figure}[h!]
    \centering
    \includegraphics[angle=0, width=15cm, height=5cm]{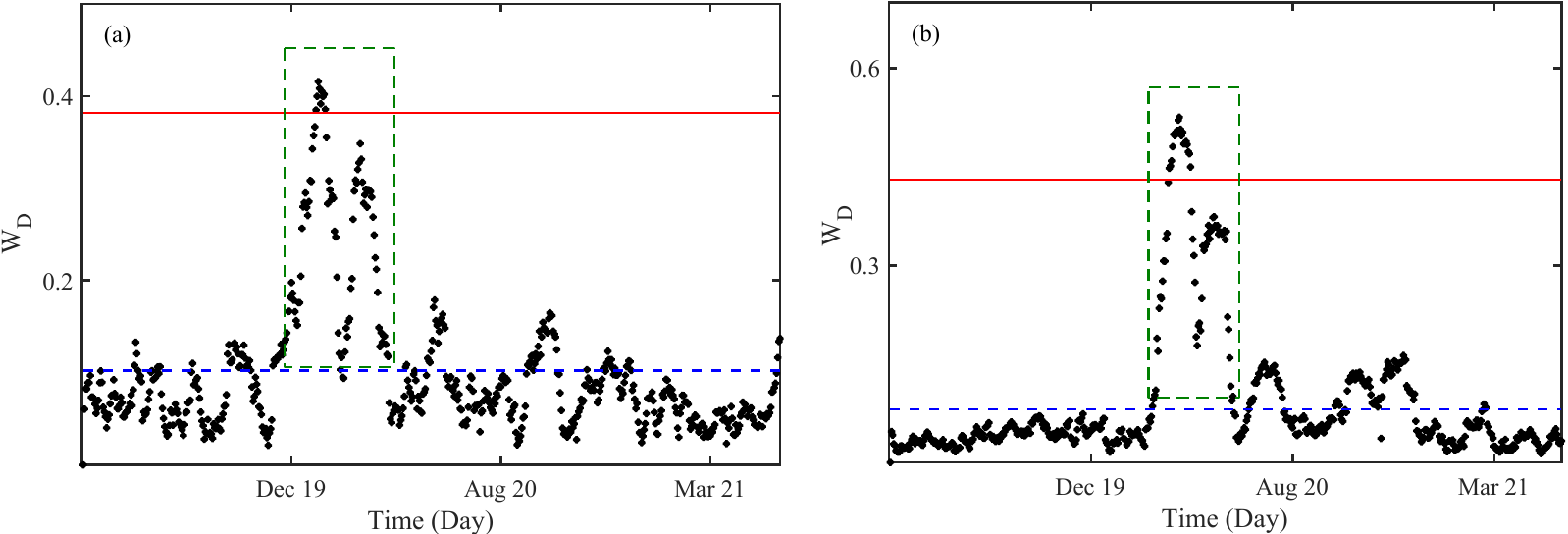}
    \caption{Plot (a) and (b) represent the Wasserstein distance ($W_D$) of Asia and Europe continents, respectively from 2019 to 2012. The spike in the $W_D$ above the threshold value coincides with the COVID-19 pandemic.}
    \label{fig:AsiaEurope_$W_D$_COVID}
\end{figure}

In order to analyze the stock market dynamics during the crash due to the COVID-19 pandemic we have estimated the $L^1~\&~L^2$ norms. Figs.~\ref{fig:AsiaEurope_COVID} (a) \& (b) represent the $L^1$ and $L^2$ norms, respectively, for Asia. A clear spike is observed in both the norms during early 2020. The abrupt rise in the norm values is consistent with the stock market crash due to the COVID-19 pandemic. The blue and red lines represent the average and threshold values, respectively. As the norms exceed the threshold value, we consider the crash in Asia due to the COVID-19 as an EE. Similarly, Figs.~\ref{fig:AsiaEurope_COVID} (c) \& (d) represent the $L^1$ and $L^2$ norms obtained for Europe. A spike in the norms is seen which exceeds the threshold value, hence, we consider the crash in Europe as an EE. We have also carried out the analysis for America and Oceania which also show similar results. The threshold of $L^1$ and $L^2$ norms for different continents during the COVID-19 is listed in Table~\ref{Table 1}.

Further, we estimate the $W_D$ to understand the stock market dynamics during the crash due to COVID-19. Fig.~\ref{fig:AsiaEurope_$W_D$_COVID} (a) \& (b) represent the $W_D$ of Asia and Europe, respectively. A clear spike in the $W_D$ is seen during early 2020 which corresponds with the actual crash time during COVID-19. As the spike in the $W_D$ is greater than the threshold, shown by the red horizontal line, we consider the crash in Asia and Europe due to COVID-19 as an EE. Analysis for Oceania also showed similar results. The threshold of $W_D$ for various continents during COVID-19 are listed in Table~\ref{Table 1}.  It was seen that during the COVID-19 pandemic, different sectors were impacted differently due to their business outlook~\cite{mazur2021covid,mahata2021modeling,rai2022sentiment}. Hence, it is important to analyze the sector-wise dynamics during the COVID-19 which is shown in the subsequent section.

\subsection{Impact of COVID-19 on different Indian sectors}

\begin{figure}[h!]
    \centering
    \includegraphics[angle=0, width=15cm, height=10cm]{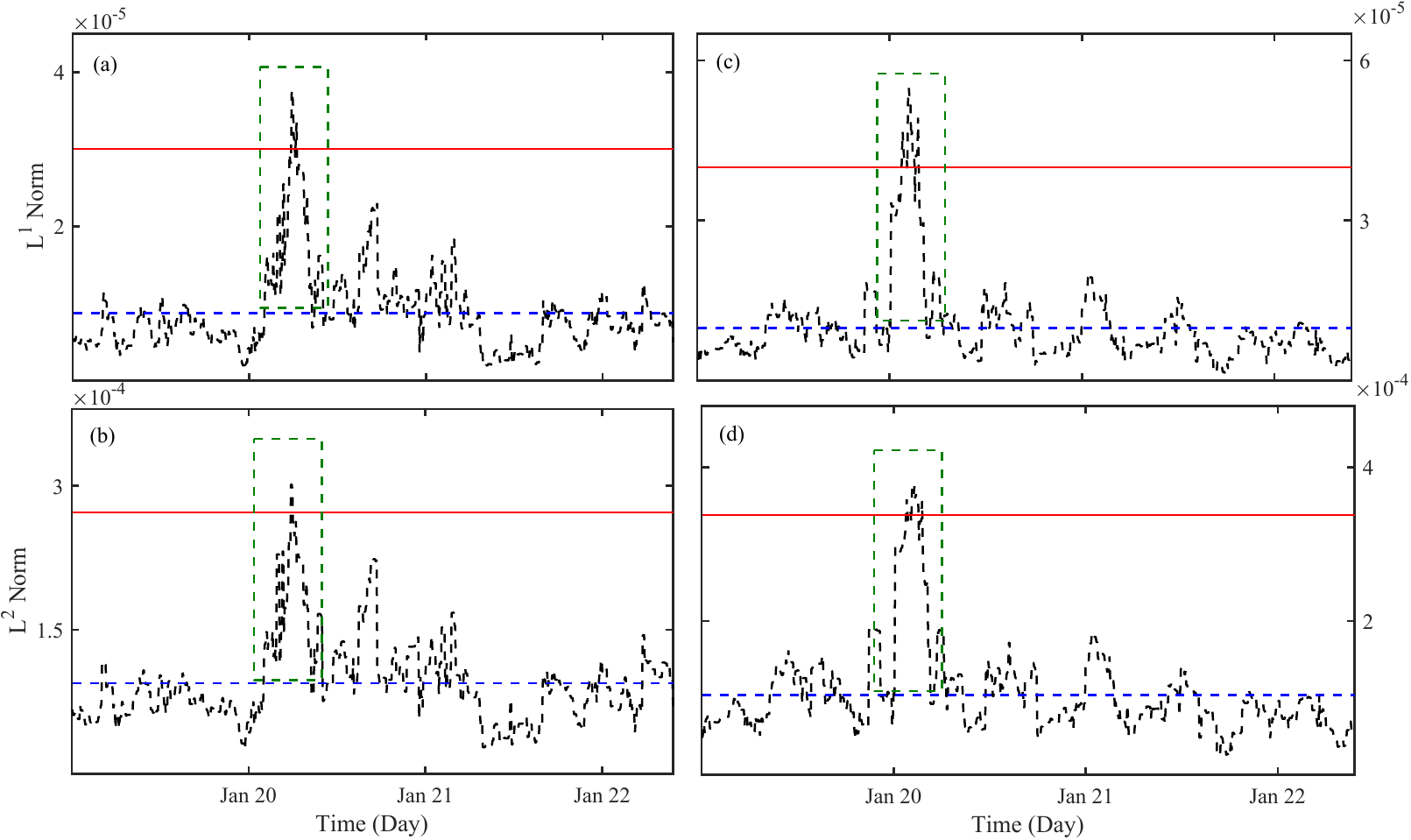}
    \caption{Plot (a) and (b) represent the $L^1$ and $L^2$ norms of the Banking sector and plot (c) and (d) represent the $L^1$ and $L^2$ norms of the Pharmaceutical sector, respectively. The abrupt rise in norm values above the threshold confirms the EE due to COVID-19.}
    \label{fig:Pharma_Bank}
\end{figure}

We analyze the impact of COVID-19 in different sectors of the Indian stock market such as Banking, Pharmaceutical, Metal, Automobiles, and fast-moving consumer goods (FMCG). For these sectors, we have taken the companies based on their respective Nifty sector indices.

\renewcommand{\arraystretch}{1.2} 
\begin{table}[h!]
\centering
\begin{tabular}{ |p{3cm}|p{2cm}|p{2cm}|p{1cm}|}
 \hline
  \multirow{2}{*}{Sectors}&\multicolumn{3}{|c|}{COVID-19} \\
\cline{2-4}
& $L^1$-norm    &$L^2$-norm& $W_D$\\
 \hline
Bank & $3.01\times 10^{-5}$ &$2.72\times 10^{-4}$ &1.48\\
 \hline
Pharmaceutical & $3.99\times 10^{-5}$ &$3.38\times 10^{-4}$ &1.86\\
 \hline
Automobile & $2.75\times 10^{-5}$ &$2.54\times 10^{-4}$ &1.17\\
 \hline
Metal & $5.54\times 10^{-5}$ &$4.14\times 10^{-4}$ &1.44\\
 \hline
FMCG & $3.00\times 10^{-5}$ &$2.46\times 10^{-4}$ &1.37\\
 \hline
\end{tabular}
\caption{Table represents the threshold of the $L^1$ norm, $L^2$ norm and $W_D$ of different sectors during the stock market crash due to COVID-19 pandemic.}
\label{sectors_table}
\end{table}
\renewcommand{\arraystretch}{1} 

\begin{figure}[h!]
    \centering
    \includegraphics[angle=0, width=15cm, height=5cm]{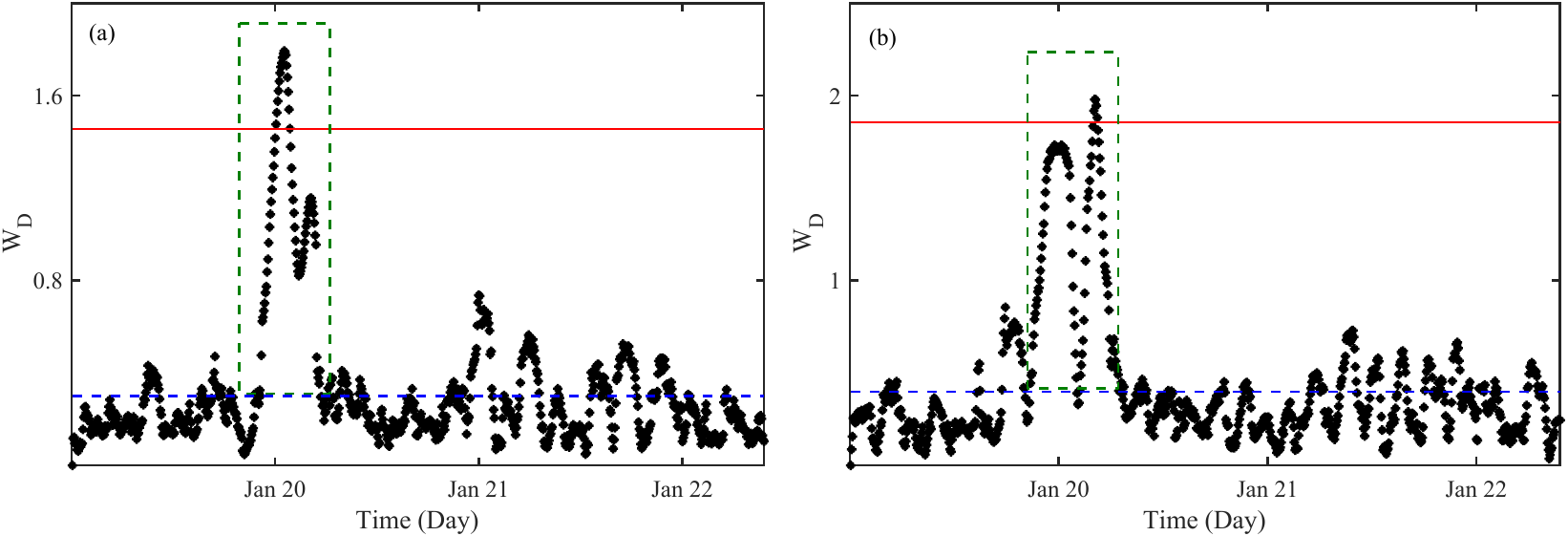}
    \caption{Plot (a) and (b) represent the Wasserstein distance ($W_D$) of the Banking and Pharmaceutical sectors, respectively from 2019 to 2012. The spike in the $W_D$ above the threshold value confirms the occurrence of EE in these sectors.}
    \label{fig:Bank_Pharma_$W_D$}
\end{figure}

Figs.~\ref{fig:Pharma_Bank} (a) \& (b) represent the $L^1$ and $L^2$ norms obtained from Banking sector. Similarly, Figs.~\ref{fig:Pharma_Bank} (c) \& (d) represent the $L^1$ and $L^2$ norms obtained from Pharmaceutical sector, respectively. A clear spike in the norms is observed during the crash in both sectors. This spike in the norms is greater than the threshold values and hence, we consider the crash in these sectors due to COVID-19 as an EE. Similar results were obtained for the automobile, Metal, and FMCG sectors. The threshold values of the $L^1$ and $L^2$ norms for these sectors are listed in Table \ref{sectors_table}. However, after the main crash, there are spikes seen in the Banking sector which are significantly high. No such spikes are observed in the Pharmaceutical. This shows that the Banking sector was in a stressful period for a longer time than the Pharmaceutical sector which recovered quickly after the crash due to its positive outlook~\cite{rai2022sentiment,mahata2021modeling}.

Figs.~\ref{fig:Bank_Pharma_$W_D$} (a) \& (b) show the $W_D$ of the Banking and Pharmaceutical sectors, respectively. The peak in the $W_D$ corresponds with the crash due to COVID-19 and also surpasses our defined threshold which is marked by the red horizontal line. Hence, the crash due to COVID-19 is considered an EE. Similar results were obtained for the automobile, Metal, and FMCG sectors. The threshold values of $W_D$ for these sectors are listed in Table \ref{sectors_table}.

\section{Conclusion}
\label{con}

Our study has proven successful in detecting continent-wise and sector-wise extreme events (EEs) in the stock market using Topological Data Analysis (TDA). The 2008 financial crisis and the crash due to the COVID-19 pandemic are identified as EEs in Asia, Europe, North-South America, and Oceania continents. A sector-wise analysis has been carried out during the COVID-19 pandemic. Earlier work has identified the crash due to COVID-19 as an EE taking the stock indices of each country separately proving it to be unsuitable for analyzing continent-wise or sector-wise. However, TDA overcomes this difficulty and we can identify continent-wise \& sector-wise EE by analyzing stock prices of $n$ stocks/indices together.

We have considered EE as those events where the spike in the $L^1$ and $L^2$ norms are greater than the defined threshold i.e., $\mu_N + 4*\sigma$. Here $\mu_N$ and $\sigma$ are the mean and the standard deviation of the norm, respectively. Also, the Wasserstein distance ($W_D$) shows an abrupt rise during the EE crossing the threshold of $\mu_W + 4*\sigma$. Here $\mu_W$ and $\sigma$ are the mean and standard deviation of the $W_D$. All the above continents show a spike in the norms and $W_D$ surpassing the thresholds during the 2008 financial crisis and the COVID-19. Hence, we consider the stock market crashes in these continents as EEs. The effect of the COVID-19 was different in different sectors, hence we have also carried out a sector-wise analysis during the COVID-19. Though the crashes in different sectors are identified as EEs, the dynamics after the crash show different impacts of COVID-19. Norm values in the Banking sector show significant spikes comparable to the main spike but no significant spikes are observed in the norm value of the Pharmaceutical sector. This shows that the Banking sector was under stress even after the crash had occurred and remained volatile.

Our results are consistent with the empirical mode decomposition-based Hilbert-Huang transform technique used for the identification of EEs by Mahata et al.~\cite{mahata2021characteristics} and Anish et al.~\cite{rai2023detection}. Hence, TDA is capable of identifying EEs in the stock market. In the future, TDA may be applied with machine learning techniques to understand the market dynamics and also may help in the prediction of stock prices.

\section{Acknowledgements}
We would like to acknowledge NIT Sikkim for allocating doctoral fellowship to Anish Rai, Buddha Nath Sharma and SR Luwang. We also like to acknowledge the inputs provided by Kundan Mukhia.

\bibliographystyle{elsarticle-num-names} 
\bibliography{ref}

\begin{thebibliography}{47}
\expandafter\ifx\csname natexlab\endcsname\relax\def\natexlab#1{#1}\fi
\providecommand{\url}[1]{\texttt{#1}}
\providecommand{\href}[2]{#2}
\providecommand{\path}[1]{#1}
\providecommand{\DOIprefix}{doi:}
\providecommand{\ArXivprefix}{arXiv:}
\providecommand{\URLprefix}{URL: }
\providecommand{\Pubmedprefix}{pmid:}
\providecommand{\doi}[1]{\href{http://dx.doi.org/#1}{\path{#1}}}
\providecommand{\Pubmed}[1]{\href{pmid:#1}{\path{#1}}}
\providecommand{\bibinfo}[2]{#2}
\ifx\xfnm\relax \def\xfnm[#1]{\unskip,\space#1}\fi
\bibitem[{Sornette(2003)}]{sornette2003critical}
\bibinfo{author}{D.~Sornette},
\newblock \bibinfo{title}{Critical market crashes},
\newblock \bibinfo{journal}{Physics reports} \bibinfo{volume}{378} (\bibinfo{year}{2003}) \bibinfo{pages}{1--98}.
\bibitem[{Vandewalle et~al.(1998)Vandewalle, Boveroux, Minguet, and Ausloos}]{vandewalle1998crash}
\bibinfo{author}{N.~Vandewalle}, \bibinfo{author}{P.~Boveroux}, \bibinfo{author}{A.~Minguet}, \bibinfo{author}{M.~Ausloos},
\newblock \bibinfo{title}{The crash of october 1987 seen as a phase transition: amplitude and universality},
\newblock \bibinfo{journal}{Physica A: Statistical Mechanics and its Applications} \bibinfo{volume}{255} (\bibinfo{year}{1998}) \bibinfo{pages}{201--210}.
\bibitem[{Choi and Douady(2012)}]{choi2012financial}
\bibinfo{author}{Y.~Choi}, \bibinfo{author}{R.~Douady},
\newblock \bibinfo{title}{Financial crisis dynamics: attempt to define a market instability indicator},
\newblock \bibinfo{journal}{Quantitative Finance} \bibinfo{volume}{12} (\bibinfo{year}{2012}) \bibinfo{pages}{1351--1365}.
\bibitem[{Mazur et~al.(2021)Mazur, Dang, and Vega}]{mazur2021covid}
\bibinfo{author}{M.~Mazur}, \bibinfo{author}{M.~Dang}, \bibinfo{author}{M.~Vega},
\newblock \bibinfo{title}{Covid-19 and the march 2020 stock market crash. evidence from s\&p1500},
\newblock \bibinfo{journal}{Finance research letters} \bibinfo{volume}{38} (\bibinfo{year}{2021}) \bibinfo{pages}{101690}.
\bibitem[{Rai et~al.(2022)Rai, Mahata, Nurujjaman, and Prakash}]{rai2022statistical}
\bibinfo{author}{A.~Rai}, \bibinfo{author}{A.~Mahata}, \bibinfo{author}{M.~Nurujjaman}, \bibinfo{author}{O.~Prakash},
\newblock \bibinfo{title}{Statistical properties of the aftershocks of stock market crashes revisited: Analysis based on the 1987 crash, financial-crisis-2008 and covid-19 pandemic},
\newblock \bibinfo{journal}{International Journal of Modern Physics C} \bibinfo{volume}{33} (\bibinfo{year}{2022}) \bibinfo{pages}{2250019}.
\bibitem[{Chen and Huang(2018)}]{chen2018panic}
\bibinfo{author}{D.-H. Chen}, \bibinfo{author}{H.-L. Huang},
\newblock \bibinfo{title}{Panic, slash, or crash—do black swans flap in stock markets?},
\newblock \bibinfo{journal}{Physica A: Statistical Mechanics and its Applications} \bibinfo{volume}{492} (\bibinfo{year}{2018}) \bibinfo{pages}{1642--1663}.
\bibitem[{Kaizoji and Sornette(2010)}]{kaizoji2010bubbles}
\bibinfo{author}{T.~Kaizoji}, \bibinfo{author}{D.~Sornette},
\newblock \bibinfo{title}{Bubbles and crashes},
\newblock \bibinfo{journal}{Encyclopedia of Quantitative Finance}  (\bibinfo{year}{2010}).
\bibitem[{Rabindrajit~Luwang et~al.(2024)Rabindrajit~Luwang, Rai, Nurujjaman, Prakash, and Hens}]{rabindrajit2024high}
\bibinfo{author}{S.~Rabindrajit~Luwang}, \bibinfo{author}{A.~Rai}, \bibinfo{author}{M.~Nurujjaman}, \bibinfo{author}{O.~Prakash}, \bibinfo{author}{C.~Hens},
\newblock \bibinfo{title}{High-frequency stock market order transitions during the us--china trade war 2018: A discrete-time markov chain analysis},
\newblock \bibinfo{journal}{Chaos: An Interdisciplinary Journal of Nonlinear Science} \bibinfo{volume}{34} (\bibinfo{year}{2024}).
\bibitem[{Albeverio et~al.(2006)Albeverio, Jentsch, and Kantz}]{albeverio2006extreme}
\bibinfo{author}{S.~Albeverio}, \bibinfo{author}{V.~Jentsch}, \bibinfo{author}{H.~Kantz}, \bibinfo{title}{Extreme events in nature and society}, \bibinfo{publisher}{Springer Science \& Business Media}, \bibinfo{year}{2006}.
\bibitem[{Mahata et~al.(2021)Mahata, Rai, Nurujjaman, Prakash, and Prasad~Bal}]{mahata2021characteristics}
\bibinfo{author}{A.~Mahata}, \bibinfo{author}{A.~Rai}, \bibinfo{author}{M.~Nurujjaman}, \bibinfo{author}{O.~Prakash}, \bibinfo{author}{D.~Prasad~Bal},
\newblock \bibinfo{title}{Characteristics of 2020 stock market crash: The covid-19 induced extreme event},
\newblock \bibinfo{journal}{Chaos: An Interdisciplinary Journal of Nonlinear Science} \bibinfo{volume}{31} (\bibinfo{year}{2021}).
\bibitem[{Raymond et~al.(2020)Raymond, Horton, Zscheischler, Martius, AghaKouchak, Balch, Bowen, Camargo, Hess, Kornhuber et~al.}]{raymond2020understanding}
\bibinfo{author}{C.~Raymond}, \bibinfo{author}{R.~M. Horton}, \bibinfo{author}{J.~Zscheischler}, \bibinfo{author}{O.~Martius}, \bibinfo{author}{A.~AghaKouchak}, \bibinfo{author}{J.~Balch}, \bibinfo{author}{S.~G. Bowen}, \bibinfo{author}{S.~J. Camargo}, \bibinfo{author}{J.~Hess}, \bibinfo{author}{K.~Kornhuber}, et~al.,
\newblock \bibinfo{title}{Understanding and managing connected extreme events},
\newblock \bibinfo{journal}{Nature climate change} \bibinfo{volume}{10} (\bibinfo{year}{2020}) \bibinfo{pages}{611--621}.
\bibitem[{Korup and Clague(2009)}]{korup2009natural}
\bibinfo{author}{O.~Korup}, \bibinfo{author}{J.~J. Clague},
\newblock \bibinfo{title}{Natural hazards, extreme events, and mountain topography},
\newblock \bibinfo{journal}{Quaternary Science Reviews} \bibinfo{volume}{28} (\bibinfo{year}{2009}) \bibinfo{pages}{977--990}.
\bibitem[{Rai et~al.(2023)Rai, Luwang, Nurujjaman, Hens, Kuila, and Debnath}]{rai2023detection}
\bibinfo{author}{A.~Rai}, \bibinfo{author}{S.~R. Luwang}, \bibinfo{author}{M.~Nurujjaman}, \bibinfo{author}{C.~Hens}, \bibinfo{author}{P.~Kuila}, \bibinfo{author}{K.~Debnath},
\newblock \bibinfo{title}{Detection and forecasting of extreme events in stock price triggered by fundamental, technical, and external factors},
\newblock \bibinfo{journal}{Chaos, Solitons \& Fractals} \bibinfo{volume}{173} (\bibinfo{year}{2023}) \bibinfo{pages}{113716}.
\bibitem[{Mahata et~al.(2020)Mahata, Bal, and Nurujjaman}]{mahata2020identification}
\bibinfo{author}{A.~Mahata}, \bibinfo{author}{D.~P. Bal}, \bibinfo{author}{M.~Nurujjaman},
\newblock \bibinfo{title}{Identification of short-term and long-term time scales in stock markets and effect of structural break},
\newblock \bibinfo{journal}{Physica A: Statistical Mechanics and its Applications} \bibinfo{volume}{545} (\bibinfo{year}{2020}) \bibinfo{pages}{123612}.
\bibitem[{Mahata and Nurujjaman(2020)}]{mahata2020time}
\bibinfo{author}{A.~Mahata}, \bibinfo{author}{M.~Nurujjaman},
\newblock \bibinfo{title}{Time scales and characteristics of stock markets in different investment horizons},
\newblock \bibinfo{journal}{Frontiers in Physics} \bibinfo{volume}{8} (\bibinfo{year}{2020}) \bibinfo{pages}{590623}.
\bibitem[{Carlsson(2009)}]{Carlsson2009TopologyAD}
\bibinfo{author}{G.~E. Carlsson},
\newblock \bibinfo{title}{Topology and data},
\newblock \bibinfo{journal}{Bulletin of the American Mathematical Society} \bibinfo{volume}{46} (\bibinfo{year}{2009}) \bibinfo{pages}{255--308}. \URLprefix \url{https://api.semanticscholar.org/CorpusID:1472609}.
\bibitem[{Dey and Wang(2022)}]{Dey_Wang_2022}
\bibinfo{author}{T.~K. Dey}, \bibinfo{author}{Y.~Wang}, \bibinfo{title}{Computational Topology for Data Analysis}, \bibinfo{publisher}{Cambridge University Press}, \bibinfo{year}{2022}.
\bibitem[{Kulkarni et~al.(2023)Kulkarni, Pharasi, Vijayaraghavan, Kumar, Chakraborti, and Samal}]{kulkarni2023investigation}
\bibinfo{author}{S.~Kulkarni}, \bibinfo{author}{H.~K. Pharasi}, \bibinfo{author}{S.~Vijayaraghavan}, \bibinfo{author}{S.~Kumar}, \bibinfo{author}{A.~Chakraborti}, \bibinfo{author}{A.~Samal},
\newblock \bibinfo{title}{Investigation of indian stock markets using topological data analysis and geometry-inspired network measures},
\newblock \bibinfo{journal}{arXiv preprint arXiv:2311.17016}  (\bibinfo{year}{2023}).
\bibitem[{Kram{\'a}r et~al.(2013)Kram{\'a}r, Goullet, Kondic, and Mischaikow}]{kramar2013persistence}
\bibinfo{author}{M.~Kram{\'a}r}, \bibinfo{author}{A.~Goullet}, \bibinfo{author}{L.~Kondic}, \bibinfo{author}{K.~Mischaikow},
\newblock \bibinfo{title}{Persistence of force networks in compressed granular media},
\newblock \bibinfo{journal}{Physical Review E} \bibinfo{volume}{87} (\bibinfo{year}{2013}) \bibinfo{pages}{042207}.
\bibitem[{Nakamura et~al.(2015)Nakamura, Hiraoka, Hirata, Escolar, and Nishiura}]{nakamura2015persistent}
\bibinfo{author}{T.~Nakamura}, \bibinfo{author}{Y.~Hiraoka}, \bibinfo{author}{A.~Hirata}, \bibinfo{author}{E.~G. Escolar}, \bibinfo{author}{Y.~Nishiura},
\newblock \bibinfo{title}{Persistent homology and many-body atomic structure for medium-range order in the glass},
\newblock \bibinfo{journal}{Nanotechnology} \bibinfo{volume}{26} (\bibinfo{year}{2015}) \bibinfo{pages}{304001}.
\bibitem[{Turner et~al.(2014)Turner, Mukherjee, and Boyer}]{turner2014persistent}
\bibinfo{author}{K.~Turner}, \bibinfo{author}{S.~Mukherjee}, \bibinfo{author}{D.~M. Boyer},
\newblock \bibinfo{title}{Persistent homology transform for modeling shapes and surfaces},
\newblock \bibinfo{journal}{Information and Inference: A Journal of the IMA} \bibinfo{volume}{3} (\bibinfo{year}{2014}) \bibinfo{pages}{310--344}.
\bibitem[{Seversky et~al.(2016)Seversky, Davis, and Berger}]{seversky2016time}
\bibinfo{author}{L.~M. Seversky}, \bibinfo{author}{S.~Davis}, \bibinfo{author}{M.~Berger},
\newblock \bibinfo{title}{On time-series topological data analysis: New data and opportunities},
\newblock in: \bibinfo{booktitle}{Proceedings of the IEEE conference on computer vision and pattern recognition workshops}, \bibinfo{year}{2016}, pp. \bibinfo{pages}{59--67}.
\bibitem[{Yao et~al.(2009)Yao, Sun, Huang, Bowman, Singh, Lesnick, Guibas, Pande, and Carlsson}]{yao2009topological}
\bibinfo{author}{Y.~Yao}, \bibinfo{author}{J.~Sun}, \bibinfo{author}{X.~Huang}, \bibinfo{author}{G.~R. Bowman}, \bibinfo{author}{G.~Singh}, \bibinfo{author}{M.~Lesnick}, \bibinfo{author}{L.~J. Guibas}, \bibinfo{author}{V.~S. Pande}, \bibinfo{author}{G.~Carlsson},
\newblock \bibinfo{title}{Topological methods for exploring low-density states in biomolecular folding pathways},
\newblock \bibinfo{journal}{The Journal of chemical physics} \bibinfo{volume}{130} (\bibinfo{year}{2009}).
\bibitem[{Nicolau et~al.(2011)Nicolau, Levine, and Carlsson}]{nicolau2011topology}
\bibinfo{author}{M.~Nicolau}, \bibinfo{author}{A.~J. Levine}, \bibinfo{author}{G.~Carlsson},
\newblock \bibinfo{title}{Topology based data analysis identifies a subgroup of breast cancers with a unique mutational profile and excellent survival},
\newblock \bibinfo{journal}{Proceedings of the National Academy of Sciences} \bibinfo{volume}{108} (\bibinfo{year}{2011}) \bibinfo{pages}{7265--7270}.
\bibitem[{Lee et~al.(2017)Lee, Barthel, D{\l}otko, Moosavi, Hess, and Smit}]{lee2017quantifying}
\bibinfo{author}{Y.~Lee}, \bibinfo{author}{S.~D. Barthel}, \bibinfo{author}{P.~D{\l}otko}, \bibinfo{author}{S.~M. Moosavi}, \bibinfo{author}{K.~Hess}, \bibinfo{author}{B.~Smit},
\newblock \bibinfo{title}{Quantifying similarity of pore-geometry in nanoporous materials},
\newblock \bibinfo{journal}{Nature communications} \bibinfo{volume}{8} (\bibinfo{year}{2017}) \bibinfo{pages}{15396}.
\bibitem[{De~Silva et~al.(2007)De~Silva, Ghrist et~al.}]{de2007homological}
\bibinfo{author}{V.~De~Silva}, \bibinfo{author}{R.~Ghrist}, et~al.,
\newblock \bibinfo{title}{Homological sensor networks},
\newblock \bibinfo{journal}{Notices of the American mathematical society} \bibinfo{volume}{54} (\bibinfo{year}{2007}).
\bibitem[{Syed~Musa et~al.(2021)Syed~Musa, Md~Noorani, Abdul~Razak, Ismail, Alias, and Hussain}]{syed2021using}
\bibinfo{author}{S.~M.~S. Syed~Musa}, \bibinfo{author}{M.~S. Md~Noorani}, \bibinfo{author}{F.~Abdul~Razak}, \bibinfo{author}{M.~Ismail}, \bibinfo{author}{M.~A. Alias}, \bibinfo{author}{S.~I. Hussain},
\newblock \bibinfo{title}{Using persistent homology as preprocessing of early warning signals for critical transition in flood},
\newblock \bibinfo{journal}{Scientific Reports} \bibinfo{volume}{11} (\bibinfo{year}{2021}) \bibinfo{pages}{7234}.
\bibitem[{Gidea(2017)}]{gidea2017topology}
\bibinfo{author}{M.~Gidea},
\newblock \bibinfo{title}{Topology data analysis of critical transitions in financial networks},
\newblock \bibinfo{journal}{arXiv preprint arXiv:1701.06081}  (\bibinfo{year}{2017}).
\bibitem[{Gidea and Katz(2018)}]{gidea2018topological}
\bibinfo{author}{M.~Gidea}, \bibinfo{author}{Y.~Katz},
\newblock \bibinfo{title}{Topological data analysis of financial time series: Landscapes of crashes},
\newblock \bibinfo{journal}{Physica A: Statistical Mechanics and its Applications} \bibinfo{volume}{491} (\bibinfo{year}{2018}) \bibinfo{pages}{820--834}.
\bibitem[{Aguilar and Ensor(2020)}]{aguilar2020topology}
\bibinfo{author}{A.~Aguilar}, \bibinfo{author}{K.~Ensor},
\newblock \bibinfo{title}{Topology data analysis using mean persistence landscapes in financial crashes},
\newblock \bibinfo{journal}{Journal of Mathematical Finance} \bibinfo{volume}{10} (\bibinfo{year}{2020}) \bibinfo{pages}{648--678}.
\bibitem[{Guo et~al.(2020)Guo, Xia, An, Zhang, Sun, and Zhao}]{guo2020empirical}
\bibinfo{author}{H.~Guo}, \bibinfo{author}{S.~Xia}, \bibinfo{author}{Q.~An}, \bibinfo{author}{X.~Zhang}, \bibinfo{author}{W.~Sun}, \bibinfo{author}{X.~Zhao},
\newblock \bibinfo{title}{Empirical study of financial crises based on topological data analysis},
\newblock \bibinfo{journal}{Physica A: Statistical Mechanics and its Applications} \bibinfo{volume}{558} (\bibinfo{year}{2020}) \bibinfo{pages}{124956}.
\bibitem[{Yen and Cheong(2021)}]{yen2021using}
\bibinfo{author}{P.~T.-W. Yen}, \bibinfo{author}{S.~A. Cheong},
\newblock \bibinfo{title}{Using topological data analysis (tda) and persistent homology to analyze the stock markets in singapore and taiwan},
\newblock \bibinfo{journal}{Frontiers in Physics} \bibinfo{volume}{9} (\bibinfo{year}{2021}) \bibinfo{pages}{572216}.
\bibitem[{Yen et~al.(2021)Yen, Xia, and Cheong}]{yen2021understanding}
\bibinfo{author}{P.~T.-W. Yen}, \bibinfo{author}{K.~Xia}, \bibinfo{author}{S.~A. Cheong},
\newblock \bibinfo{title}{Understanding changes in the topology and geometry of financial market correlations during a market crash},
\newblock \bibinfo{journal}{Entropy} \bibinfo{volume}{23} (\bibinfo{year}{2021}) \bibinfo{pages}{1211}.
\bibitem[{Goel et~al.(2020)Goel, Pasricha, and Mehra}]{goel2020topological}
\bibinfo{author}{A.~Goel}, \bibinfo{author}{P.~Pasricha}, \bibinfo{author}{A.~Mehra},
\newblock \bibinfo{title}{Topological data analysis in investment decisions},
\newblock \bibinfo{journal}{Expert Systems with Applications} \bibinfo{volume}{147} (\bibinfo{year}{2020}) \bibinfo{pages}{113222}.
\bibitem[{Majumdar and Laha(2020)}]{majumdar2020clustering}
\bibinfo{author}{S.~Majumdar}, \bibinfo{author}{A.~K. Laha},
\newblock \bibinfo{title}{Clustering and classification of time series using topological data analysis with applications to finance},
\newblock \bibinfo{journal}{Expert Systems with Applications} \bibinfo{volume}{162} (\bibinfo{year}{2020}) \bibinfo{pages}{113868}.
\bibitem[{Gidea et~al.(2020)Gidea, Goldsmith, Katz, Roldan, and Shmalo}]{gidea2020topological}
\bibinfo{author}{M.~Gidea}, \bibinfo{author}{D.~Goldsmith}, \bibinfo{author}{Y.~Katz}, \bibinfo{author}{P.~Roldan}, \bibinfo{author}{Y.~Shmalo},
\newblock \bibinfo{title}{Topological recognition of critical transitions in time series of cryptocurrencies},
\newblock \bibinfo{journal}{Physica A: Statistical mechanics and its applications} \bibinfo{volume}{548} (\bibinfo{year}{2020}) \bibinfo{pages}{123843}.
\bibitem[{Souto(2023)}]{souto2023topological}
\bibinfo{author}{H.~G. Souto},
\newblock \bibinfo{title}{Topological tail dependence: Evidence from forecasting realized volatility},
\newblock \bibinfo{journal}{The Journal of Finance and Data Science} \bibinfo{volume}{9} (\bibinfo{year}{2023}) \bibinfo{pages}{100107}.
\bibitem[{Guo et~al.(2021)Guo, Zhao, Yu, and Zhang}]{guo2021analysis}
\bibinfo{author}{H.~Guo}, \bibinfo{author}{X.~Zhao}, \bibinfo{author}{H.~Yu}, \bibinfo{author}{X.~Zhang},
\newblock \bibinfo{title}{Analysis of global stock markets’ connections with emphasis on the impact of covid-19},
\newblock \bibinfo{journal}{Physica A: Statistical Mechanics and its Applications} \bibinfo{volume}{569} (\bibinfo{year}{2021}) \bibinfo{pages}{125774}.
\bibitem[{Guo et~al.(2022)Guo, Yu, An, and Zhang}]{guo2022risk}
\bibinfo{author}{H.~Guo}, \bibinfo{author}{H.~Yu}, \bibinfo{author}{Q.~An}, \bibinfo{author}{X.~Zhang},
\newblock \bibinfo{title}{Risk analysis of china’s stock markets based on topological data structures},
\newblock \bibinfo{journal}{Procedia Computer Science} \bibinfo{volume}{202} (\bibinfo{year}{2022}) \bibinfo{pages}{203--216}.
\bibitem[{Mahata et~al.(2021)Mahata, Rai, Nurujjaman, and Prakash}]{mahata2021modeling}
\bibinfo{author}{A.~Mahata}, \bibinfo{author}{A.~Rai}, \bibinfo{author}{M.~Nurujjaman}, \bibinfo{author}{O.~Prakash},
\newblock \bibinfo{title}{Modeling and analysis of the effect of covid-19 on the stock price: V and l-shape recovery},
\newblock \bibinfo{journal}{Physica A: Statistical Mechanics and its Applications} \bibinfo{volume}{574} (\bibinfo{year}{2021}) \bibinfo{pages}{126008}.
\bibitem[{Gold et~al.(2023)Gold, Karabina, and Motta}]{gold2023algorithm}
\bibinfo{author}{D.~Gold}, \bibinfo{author}{K.~Karabina}, \bibinfo{author}{F.~C. Motta},
\newblock \bibinfo{title}{An algorithm for persistent homology computation using homomorphic encryption},
\newblock \bibinfo{journal}{arXiv preprint arXiv:2307.01923}  (\bibinfo{year}{2023}).
\bibitem[{Akingbade et~al.(2024)Akingbade, Gidea, Manzi, and Nateghi}]{akingbade2024topological}
\bibinfo{author}{S.~W. Akingbade}, \bibinfo{author}{M.~Gidea}, \bibinfo{author}{M.~Manzi}, \bibinfo{author}{V.~Nateghi},
\newblock \bibinfo{title}{Why topological data analysis detects financial bubbles?},
\newblock \bibinfo{journal}{Communications in Nonlinear Science and Numerical Simulation} \bibinfo{volume}{128} (\bibinfo{year}{2024}) \bibinfo{pages}{107665}.
\bibitem[{Chazal and Michel(2021)}]{chazal2021introduction}
\bibinfo{author}{F.~Chazal}, \bibinfo{author}{B.~Michel},
\newblock \bibinfo{title}{An introduction to topological data analysis: fundamental and practical aspects for data scientists},
\newblock \bibinfo{journal}{Frontiers in artificial intelligence} \bibinfo{volume}{4} (\bibinfo{year}{2021}) \bibinfo{pages}{108}.
\bibitem[{Cohen-Steiner et~al.(2005)Cohen-Steiner, Edelsbrunner, and Harer}]{cohen2005stability}
\bibinfo{author}{D.~Cohen-Steiner}, \bibinfo{author}{H.~Edelsbrunner}, \bibinfo{author}{J.~Harer},
\newblock \bibinfo{title}{Stability of persistence diagrams},
\newblock in: \bibinfo{booktitle}{Proceedings of the twenty-first annual symposium on Computational geometry}, \bibinfo{year}{2005}, pp. \bibinfo{pages}{263--271}.
\bibitem[{Bubenik(2015)}]{JMLR:v16:bubenik15a}
\bibinfo{author}{P.~Bubenik},
\newblock \bibinfo{title}{Statistical topological data analysis using persistence landscapes},
\newblock \bibinfo{journal}{Journal of Machine Learning Research} \bibinfo{volume}{16} (\bibinfo{year}{2015}) \bibinfo{pages}{77--102}. \URLprefix \url{http://jmlr.org/papers/v16/bubenik15a.html}.
\bibitem[{yahoo(2024)}]{Yahoo}
yahoo, \bibinfo{title}{Yahoo finance}, \bibinfo{howpublished}{\url{https://finance.yahoo.com/}}, \bibinfo{year}{2024}.
\bibitem[{Rai et~al.(2022)Rai, Mahata, Nurujjaman, Majhi, and Debnath}]{rai2022sentiment}
\bibinfo{author}{A.~Rai}, \bibinfo{author}{A.~Mahata}, \bibinfo{author}{M.~Nurujjaman}, \bibinfo{author}{S.~Majhi}, \bibinfo{author}{K.~Debnath},
\newblock \bibinfo{title}{A sentiment-based modeling and analysis of stock price during the covid-19: U-and swoosh-shaped recovery},
\newblock \bibinfo{journal}{Physica A: Statistical Mechanics and its Applications} \bibinfo{volume}{592} (\bibinfo{year}{2022}) \bibinfo{pages}{126810}.

\end{thebibliography}
\end{document}